\pdfoutput=1
\documentclass[runningheads]{llncs}

\usepackage[T1]{fontenc}
\usepackage{amsmath}
\usepackage{newtxtext}
\usepackage[varvw]{newtxmath}
\usepackage{graphicx}
\usepackage{booktabs}
\usepackage{placeins}
\usepackage{cite}
\usepackage{color}
\usepackage{hyperref}

\begin{document}

\title{Preserving Orientation and Transition Order in Transition Network Analysis: A Directed Topological Proof of Concept}
\titlerunning{Directed Topological Proof of Concept in TNA}

\author{Koichi Yasutake\inst{1} \and Hitoshi Inoue\inst{2}}
\authorrunning{K. Yasutake and H. Inoue}

\institute{Hiroshima University, Japan\\
\email{ystake@hiroshima-u.ac.jp}
\and
Nakamura Gakuen University, Japan\\
\email{jinoue@nakamura-u.ac.jp}}

\maketitle

\begin{abstract}
Transition Network Analysis (TNA), implemented in R packages such as \texttt{tna} and \texttt{Nestimate}, models learning processes as directed Markov transition probabilities. While \texttt{Nestimate} also provides sequence-level higher-order and pathway models, existing clique-based topological tools lose directional structure in two ways: \texttt{tna::cliques()} returns unordered, mutually connected subsets without homology or convergence--divergence asymmetry, whereas undirected persistent homology in \texttt{Nestimate} symmetrizes the transition matrix prior to constructing clique complexes, erasing edge orientation and the orientation order of directed triangles. We evaluate four directed topological constructions on the \texttt{tna} benchmark dataset \texttt{group\_regulation} (9 collaborative regulation states): Dowker sink and source complexes, directed flag complexes ($dFl$), and standard (regular) GLMY path homology, compared against row-wise entry-permutation and degree-preserving null models ($B = 20{,}000$). Across $\tau \in [0.20, 0.28]$, $\beta_1^{dFl}$ exceeds the max-symmetrized clique complex $\beta_1^{\mathrm{sym}}$ by exactly $1$; for $\tau \in (0.239087, 0.290401]$ (represented by $\tau = 0.26$), $\beta_1^{\mathrm{sym}} = 0$ whereas $dFl$ preserves $\beta_1^{dFl} = 1$, corresponding to the reciprocal \texttt{consensus} $\leftrightarrow$ \texttt{plan} loop (weights $0.3958$ and $0.2904$). Across $\tau \in (0.146825, 0.172081]$, $dFl$ also leaves the directed 3-cycle \texttt{\{consensus, coregulate, emotion\}} unfilled while max-symmetrization fills it. Regular GLMY path homology yields $\beta_1^{\mathrm{path}} = 0$ across all thresholds, showing that all directed cycles are homologous to zero in the regular path complex (with a lower-tail $p = 0.012$ for the $\beta_1^{\mathrm{path}}$ sum over $\tau \in [0.14, 0.30]$ under row-wise entry permutation, reflecting strong in-degree concentration). Dowker analysis further reveals a significant sink--source asymmetry: 20--21 sink 2-simplices versus 0 source 2-simplices across $\tau \in [0.20, 0.28]$ ($p = 0.002$ at $\tau = 0.26$), driven by six behaviors converging onto \texttt{consensus} ($\mathrm{InStrength} = 2.667$). Although \texttt{tna::cliques(size = 2)} and \texttt{InStrength} detect the reciprocal pair and incoming weight separately, directed complexes express both as oriented 1-homology classes and higher-order simplices over the same directed graph.

\keywords{Transition Network Analysis \and Directed Topological Data Analysis \and Dowker Complex \and Directed Flag Complex \and GLMY Path Homology \and Collaborative Learning.}
\end{abstract}

\section{Introduction}
Transition Network Analysis (TNA) has emerged as a prominent framework in learning analytics for modeling sequential learning behaviors as first-order Markov transition networks \cite{saqr2025tna,tikka2025tna_pkg}. Its companion package, \texttt{Nestimate}, extends this workflow with sequence-level higher-order Markov modeling, pathway anomaly detection, hypergraphs, simplicial complexes, and persistent homology \cite{saqr2026nestimate}. While TNA estimates an inherently \emph{asymmetric} transition probability matrix $W \in [0,1]^{n \times n}$, its clique-based topological tools lose directional structure in two distinct ways: \texttt{tna::cliques()} (by default) requires mutual edges and returns unordered, mutually connected subsets without distinguishing orientation patterns, homology, or convergence versus divergence, whereas undirected persistent homology in \texttt{Nestimate} relies on symmetrizing the weighted matrix into an undirected graph prior to constructing clique complexes.

As shown in recent work on co-occurrence data in learning analytics \cite{yasutake2026beyond}, topological constructions should match the structural properties of the underlying data. Here, we use \emph{transition order} (or \emph{orientation order}) to refer to the edge-orientation pattern within ordered simplices of the aggregated transition graph (distinct from Markov chain order or raw sequence order). For clique-based topological analyses of directed transition data, max-symmetrization introduces two fundamental distortions:
\begin{enumerate}
  \item \textbf{Loss of orientation}: Forward and backward transitions ($u \to v$ and $v \to u$) collapse into a single undirected edge, erasing reciprocal feedback loops from homology and making convergence (many-to-one) indistinguishable from divergence (one-to-many).
  \item \textbf{Loss of orientation order in triangles}: Undirected clique complexes fill any pairwise triangle $\{u,v,w\}$ into a 2-simplex regardless of whether the edges form a transitive feed-forward progression or a non-transitive cycle.
\end{enumerate}
In this proof-of-concept study, we compare standard \texttt{tna}/\texttt{Nestimate} baselines and a max-symmetrized clique complex against four directed topological constructions---Dowker sink complexes, Dowker source complexes \cite{dowker1952homology,chowdhury2018dowker}, directed flag complexes \cite{lutgehetmann2020flagser}, and GLMY path homology \cite{grigoryan2012path,chowdhury2018pph}---on the \texttt{group\_regulation} benchmark dataset. While \texttt{tna::cliques(size = 2)} extracts mutual pairs and node-level \texttt{InStrength} reflects incoming weight, directed complexes represent reciprocal feedback as oriented 1-homology classes and sink convergence as 2-simplices in the common language of complexes over the same directed graph.

\section{Directed Topological Constructions and Baselines}
Let $W = (w_{ij})$ be the $n \times n$ row-stochastic transition probability matrix estimated by TNA. For a filtration threshold $\tau \in [0.02, 0.30]$, we define the binary directed adjacency matrix $A^{(\tau)}_{ij} = \mathbb{I}(w_{ij} \ge \tau)$ with $i \neq j$, yielding a directed graph $G_\tau = (V, E_\tau)$.

\subsubsection{Undirected Clique Baselines.}
We construct the max-symmetrized graph $U_\tau$ with adjacency entries $U^{(\tau)}_{ij} = \max(A^{(\tau)}_{ij}, A^{(\tau)}_{ji})$ and form its clique complex $\mathrm{Cl}(U_\tau)$, whose $k$-simplices are all $(k+1)$-cliques in $U_\tau$ (yielding $\beta_1^{\mathrm{sym}}$). Separately, we run \texttt{tna::cliques()} (which requires mutual edges by default), \texttt{Nestimate::\allowbreak betti\_numbers()}, and \texttt{Nestimate::\allowbreak persistent\_homology()} directly in R as package baselines.

\subsubsection{Dowker Sink and Source Complexes.}
Following Chowdhury and M\'emoli \cite{chowdhury2018dowker}, the Dowker sink complex $D_{\mathrm{sink}}(G_\tau)$ contains a simplex $\sigma \subseteq V$ if and only if there exists a common target $y \in V$ such that $(x, y) \in E_\tau$ for all $x \in \sigma$. Dually, the Dowker source complex $D_{\mathrm{source}}(G_\tau)$ contains $\sigma \subseteq V$ if and only if there exists a common source $x \in V$ such that $(x, y) \in E_\tau$ for all $y \in \sigma$. While Dowker duality guarantees isomorphic homology $H_*(D_{\mathrm{sink}}(G_\tau)) \cong H_*(D_{\mathrm{source}}(G_\tau))$ \cite{dowker1952homology,chowdhury2018dowker}, comparing their simplex counts reveals convergence--divergence asymmetry.

\subsubsection{Directed Flag Complex ($dFl$).}
The directed flag complex $dFl(G_\tau)$ is the ordered simplicial complex whose $k$-simplices are ordered $(k+1)$-tuples $(v_0, \dots, v_k)$ satisfying $(v_i, v_j) \in E_\tau$ for all $0 \le i < j \le k$ \cite{lutgehetmann2020flagser}. Unlike max-symmetrized clique complexes, $dFl(G_\tau)$ fills only acyclic transitive tournaments and leaves both reciprocal 2-cycles ($a \leftrightarrow b$) and non-transitive cycles of length $\ge 3$ unfilled (though an unfilled cycle can still be bounding when adjacent transitive triangles form a 2-chain boundary). Forgetting vertex ordering on each ordered simplex defines a forgetful simplicial map $dFl(G_\tau) \to \mathrm{Cl}(U_\tau)$, inducing a linear map $H_1(dFl(G_\tau)) \to H_1(\mathrm{Cl}(U_\tau))$.

\subsubsection{GLMY Path Homology.}
We also compute standard (regular) GLMY path homology \cite{grigoryan2012path,chowdhury2018pph}. Let $\mathcal{A}_p(G_\tau)$ be the vector space spanned by elementary allowed $p$-paths $(v_0, \dots, v_p)$ with $(v_i, v_{i+1}) \in E_\tau$ and $v_i \neq v_{i+1}$, quotienting out degenerate faces with consecutive repeated vertices. The $\partial$-invariant chain space is $\Omega_p(G_\tau) = \{ u \in \mathcal{A}_p \mid \partial_p u \in \mathcal{A}_{p-1} \}$, and $\beta_p^{\mathrm{path}} = \dim \ker(\partial_p|_{\Omega_p}) - \operatorname{rank}(\partial_{p+1}|_{\Omega_{p+1}})$. Because $\partial_2 e_{a,b,a} = e_{b,a} + e_{a,b} \in \mathcal{A}_1$ in the regular complex, reciprocal 2-cycles are filled ($\beta_1^{\mathrm{path}} = 0$); longer cycles can also be filled when bridged by $\Omega_2$ boundaries.

\subsubsection{Null Models.}
To test whether observed structures depart from random expectations ($B = 20{,}000$, NumPy \texttt{default\_rng(42)}), we evaluate:
(1) \emph{Row-wise entry permutation}, which shuffles off-diagonal entries independently within each row of $W$ (preserving row sums, out-weights, and self-loops, so the sink--source asymmetry test specifically evaluates sink concentration); and
(2) \emph{Degree-preserving directed rewiring} \cite{maslov2002specificity} on $G_{0.26}$ ($150$ swap attempts per replicate, preserving exact in- and out-degrees).

\subsubsection{Sequence-Level Bootstrap Resampling.}
To assess sampling stability, we draw $B = 1{,}000$ bootstrap resamples with replacement over the $2{,}000$ \texttt{group\_regulation} learner sequences (R \texttt{set.seed(42)}), re-estimating $W$ via \texttt{tna()} on each resample and thresholding primarily at $\tau = 0.26$ (with $\tau = 0.24$ for comparing the symmetrization blind spot, and $\tau = 0.16$ for evaluating the directed 3-cycle). We say that a feature \emph{persists} in a resample if its defining criterion holds in that resample's thresholded digraph: both edges of \texttt{consensus} $\leftrightarrow$ \texttt{plan} are present; sink 2-simplices outnumber source 2-simplices; the symmetrization blind spot ($\beta_1^{\mathrm{sym}} = 0$ and $\beta_1^{dFl} = 1$) holds; or the directed 3-cycle \texttt{\{consensus, coregulate, emotion\}} is present and has no ordered 2-simplex on that triple in $dFl$.

\section{Results on \texttt{group\_regulation}}
We analyze the \texttt{group\_regulation} dataset ($n = 9$ states: \texttt{adapt}, \texttt{cohesion}, \texttt{consensus}, \texttt{coregulate}, \texttt{discuss}, \texttt{emotion}, \texttt{monitor}, \texttt{plan}, \texttt{synthesis}). In R (\texttt{tna} v1.2.3 and \texttt{Nestimate} v1.0.12), \texttt{tna::cliques()} (requiring mutual edges) yields $3$ undirected 3-cliques at $\tau = 0.05$ and $0$ for $\tau \ge 0.10$ (compared with $50$ triangles in the max-symmetrized graph at $\tau = 0.05$); \texttt{Nestimate::betti\_numbers()} on the full complex yields $\beta_0 = 1, \beta_k = 0$ ($k \ge 1$); and \texttt{Nestimate}'s \texttt{persistent\_homology()} ($20$ filtration steps across $[0.4979, 0.0050]$) detects $2$ finite-lived undirected $\beta_1$ features (\texttt{b1: 2}), with $0$ surviving to the complete-graph endpoint.

\begin{figure}[t]
  \centering
  \includegraphics[width=\textwidth]{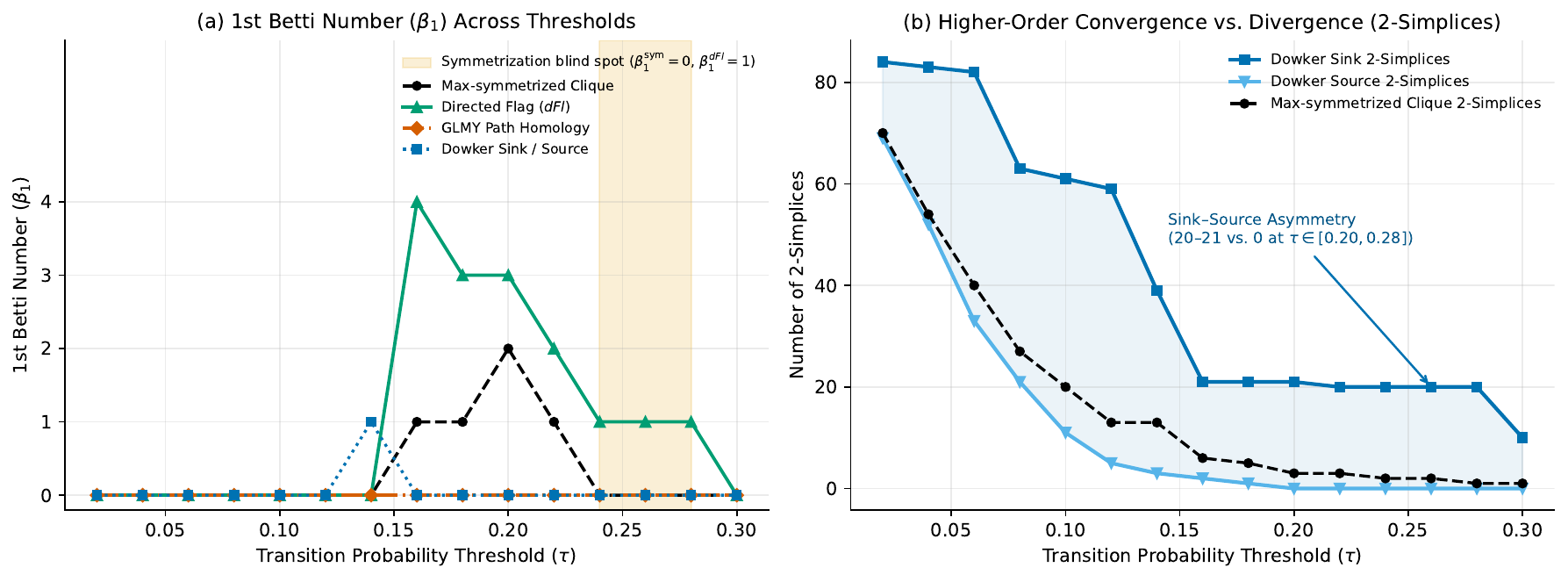}
  \caption{Filtration comparison on \texttt{group\_regulation}. (a) First Betti number $\beta_1$ across the $0.02$-step grid: across $\tau \in [0.20, 0.28]$, $\beta_1^{dFl} - \beta_1^{\mathrm{sym}} = 1$, and on the $0.02$ grid for $\tau \in [0.24, 0.28]$ max-symmetrized cliques drop to $\beta_1^{\mathrm{sym}} = 0$ while $dFl$ retains $\beta_1^{dFl} = 1$; regular GLMY path homology remains $\beta_1^{\mathrm{path}} = 0$, and Dowker complexes show a transient $\beta_1 = 1$ at $\tau = 0.14$. (b) 2-simplex counts showing extreme Dowker sink--source asymmetry.}
  \label{fig:filtration}
\end{figure}

\subsection{Reciprocal Feedback and a Directed 3-Cycle Missed by Symmetrization}
\begin{figure}[t]
  \centering
  \includegraphics[width=\textwidth]{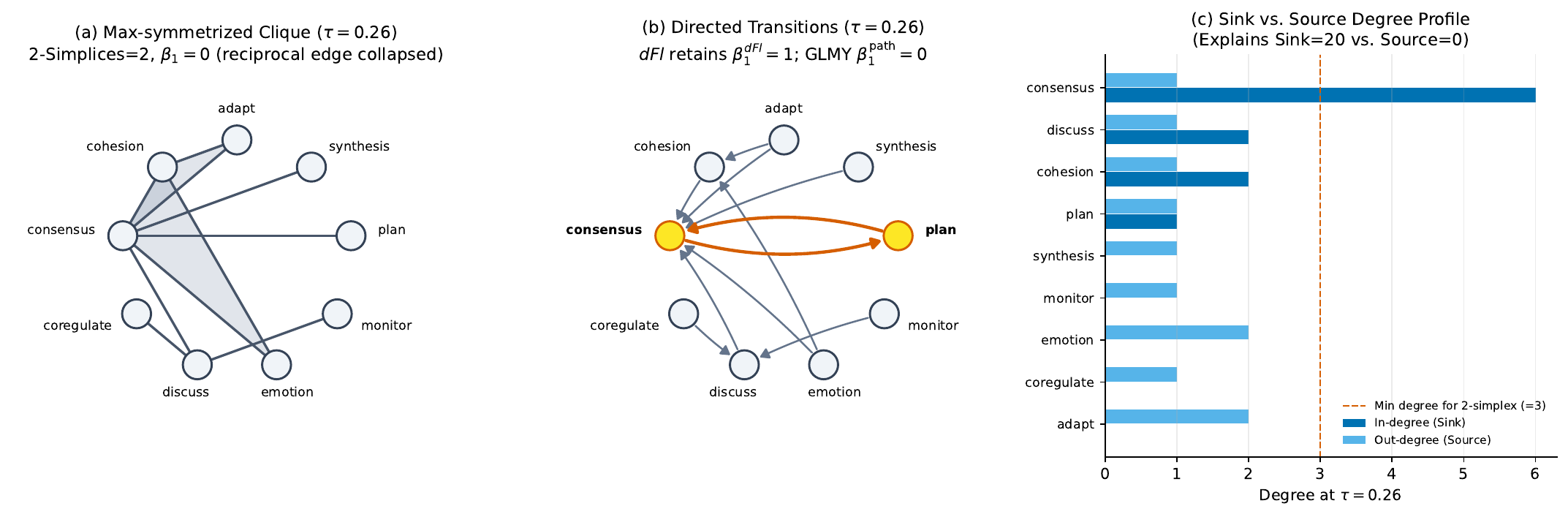}
  \caption{Mechanism at $\tau = 0.26$ (identical to $\tau = 0.24$, since $G_\tau$ is constant on $(0.239087, 0.273084]$). (a) Max-symmetrized clique complex ($\beta_1^{\mathrm{sym}} = 0$). (b) Directed network highlighting the reciprocal \texttt{consensus} $\leftrightarrow$ \texttt{plan} cycle ($\beta_1^{dFl} = 1$; $\beta_1^{\mathrm{path}} = 0$). (c) In-degree versus out-degree profile explaining the $20$-versus-$0$ sink--source gap.}
  \label{fig:mechanism}
\end{figure}

As shown in \mbox{Fig.~\ref{fig:filtration}(a)}, across $\tau \in [0.20, 0.28]$, the forgetful map $H_1(dFl(G_\tau)) \to H_1(\mathrm{Cl}(U_\tau))$ is surjective with a 1-dimensional kernel, so $\beta_1^{dFl} - \beta_1^{\mathrm{sym}} = 1$ ($3$ vs.\ $2$, $2$ vs.\ $1$, and $1$ vs.\ $0$). In particular, $\beta_1^{\mathrm{sym}} = 0$ while $\beta_1^{dFl} = 1$ holds over the continuous interval $\tau \in (0.239087, 0.290401]$, where $0.239087$ rounds the exact \texttt{coregulate} $\to$ \texttt{plan} weight ($471/1970 = 0.239086\dots$) upward into the open interval. Because $\tau = 0.24$ sits only $0.000914$ above $0.239086$ whereas $G_\tau$ is identical across $\tau \in (0.239087, 0.273084]$ ($G_{0.24} = G_{0.26}$), we use $\tau = 0.26$ as our primary interior reference threshold. Direct cycle enumeration confirms that across $\tau \in [0.20, 0.28]$ there are no simple directed cycles of length $3$ to $9$, and the sole elementary directed cycle is the reciprocal pair:
\[
  \texttt{consensus} \xrightarrow{0.3958} \texttt{plan} \xrightarrow{0.2904} \texttt{consensus},
\]
which spans the 1-dimensional kernel above, while the remaining classes at $\tau \in \{0.20, 0.22\}$ match those of $\mathrm{Cl}(U_\tau)$ (\mbox{Fig.~\ref{fig:mechanism}(a)--(b)}). The two undirected triangles present at $\tau = 0.26$---\texttt{\{adapt, cohesion, consensus\}} and \texttt{\{cohesion, consensus, emotion\}}---are both transitive in $dFl$.

At lower thresholds ($\tau \in [0.14, 0.18]$), a second reciprocal pair (\texttt{consensus} $\leftrightarrow$ \texttt{discuss}) and longer cycles appear, and Dowker complexes exhibit a transient $\beta_1 = 1$ at $\tau = 0.14$. Moreover, for $\tau \in (0.134518, 0.172081]$ (between \texttt{coregulate} $\to$ \texttt{consensus} $= 0.134518$ and \texttt{coregulate} $\to$ \texttt{emotion} $= 0.172081$), the triple \texttt{\{consensus, coregulate, emotion\}} forms a directed 3-cycle (\texttt{consensus} $\to$ \texttt{coregulate} $\to$ \texttt{emotion} $\to$ \texttt{consensus}) with no ordered 2-simplex in $dFl(G_\tau)$, whereas $\mathrm{Cl}(U_\tau)$ fills it as a 2-simplex. Its $dFl$ homology class is non-trivial across $\tau \in (0.146825, 0.172081]$, forming an independent class within the 3-dimensional kernel at $\tau = 0.16$ (and at $\tau = 0.15$, off the $0.02$ grid, $\beta_1^{\mathrm{sym}} = 0$ while $\beta_1^{dFl} = 1$ represents this directed 3-cycle without the two reciprocal pairs, which are bounding there; at $\tau = 0.14$, $\beta_1^{dFl} = 0$). Taken together, while \texttt{tna::cliques(size = 2, threshold = 0.24)} (identical at $\tau = 0.26$) extracts \texttt{\{consensus, plan\}}, max-symmetrization erases both the reciprocal loop and this directed 3-cycle from homology, whereas $dFl$ preserves them as oriented 1-homology classes. Meanwhile, regular GLMY path homology yields $\beta_1^{\mathrm{path}} = 0$ across all thresholds, indicating that all directed cycles are homologous to zero in the regular path complex. In the sequence-level bootstrap at $\tau = 0.26$ ($B = 1{,}000$, \texttt{set.seed(42)}), the \texttt{consensus} $\leftrightarrow$ \texttt{plan} loop ($100.0\%$), sink--source asymmetry ($100.0\%$), and the symmetrization blind spot ($99.0\%$, versus $52.7\%$ at $\tau = 0.24$) persist, while at $\tau = 0.16$ the directed 3-cycle \texttt{\{consensus, coregulate, emotion\}} remains unfilled (has no ordered 2-simplex on that triple in $dFl$) in $91.5\%$ of resamples. At $\tau = 0.26$, the full 11-edge graph $G_{0.26}$ matches identically in $63.5\%$ of resamples, mainly because \texttt{adapt} $\to$ \texttt{cohesion} ($0.2731$) and \texttt{coregulate} $\to$ \texttt{discuss} ($0.2736$) lie just above $0.26$ (dropping below $0.26$ in $24.1\%$ and $9.0\%$ of resamples, respectively, while \texttt{synthesis} $\to$ \texttt{adapt} at $0.2347$ enters in $6.7\%$).

\subsection{Sink--Source Asymmetry and Null Model Comparisons}
Dowker complexes uncover a sharp asymmetry invisible to undirected analyses (\mbox{Fig.~\ref{fig:filtration}(b)}, \mbox{Fig.~\ref{fig:mechanism}(c)}). Across $\tau \in [0.20, 0.28]$, the Dowker sink complex contains $20$--$21$ 2-simplices, whereas the Dowker source complex contains $0$. At $\tau = 0.26$ (and $\tau = 0.24$), six states (\texttt{cohesion} $0.4979$, \texttt{adapt} $0.4774$, \texttt{synthesis} $0.4663$, \texttt{discuss} $0.3212$, \texttt{emotion} $0.3204$, and \texttt{plan} $0.2904$) all transition strongly into \texttt{consensus} ($\mathrm{InStrength} = 2.667$), generating $\binom{6}{3} = 20$ sink 2-simplices; at $\tau = 0.20$, a 21st sink 2-simplex \texttt{\{consensus, coregulate, monitor\}} appears with \texttt{plan} as its common target.

\begin{table}[t]
  \centering
  \caption{Null model comparisons ($B = 20{,}000$; one-sided empirical $p$-values \mbox{$(1 + k)/(B + 1)$}).}
  \label{tab:null}
  \resizebox{\textwidth}{!}{%
  \begin{tabular}{llrrrrl}
    \toprule
    Null Model & Metric & Obs. & Mean & SD & $p$-value & Tail \\
    \midrule
    Row-wise Entry Perm. & Sink--Source asym.\ ($\tau = 0.20$) & 21 & 3.563 & 3.180 & 0.003 & upper \\
    Row-wise Entry Perm. & Sink--Source asym.\ ($\tau = 0.26$) & 20 & 1.720 & 2.142 & 0.002 & upper \\
    Row-wise Entry Perm. & GLMY $\beta_1^{\mathrm{path}}$ sum $[0.14, 0.30]$ & 0 & 13.070 & 6.024 & 0.012 & lower \\
    Row-wise Entry Perm. & $\beta_1^{dFl}$ sum $[0.14, 0.30]$ & 15 & 29.441 & 7.606 & 0.040 & lower \\
    Degree-Preserving & GLMY $\beta_1^{\mathrm{path}}$ ($\tau = 0.26$)$^\dagger$ & 0 & 0.576 & 0.673 & 0.523 & lower \\
    Degree-Preserving & Symmetrization blind spot ($\tau = 0.26$)$^\dagger$ & 1 & 0.324 & 0.468 & 0.324 & upper \\
    \bottomrule
  \end{tabular}}
  \\[4pt]
  {\footnotesize $^\dagger$Exact fractions across all $437$ digraphs with the observed $\tau = 0.26$ ($= \tau = 0.24$) in- and out-degree sequence: $229/437 = 0.524$ for $\beta_1^{\mathrm{path}} = 0$, and $142/437 = 0.325$ for the symmetrization blind spot.}
\end{table}

\begin{figure}[t]
  \centering
  \includegraphics[width=\textwidth]{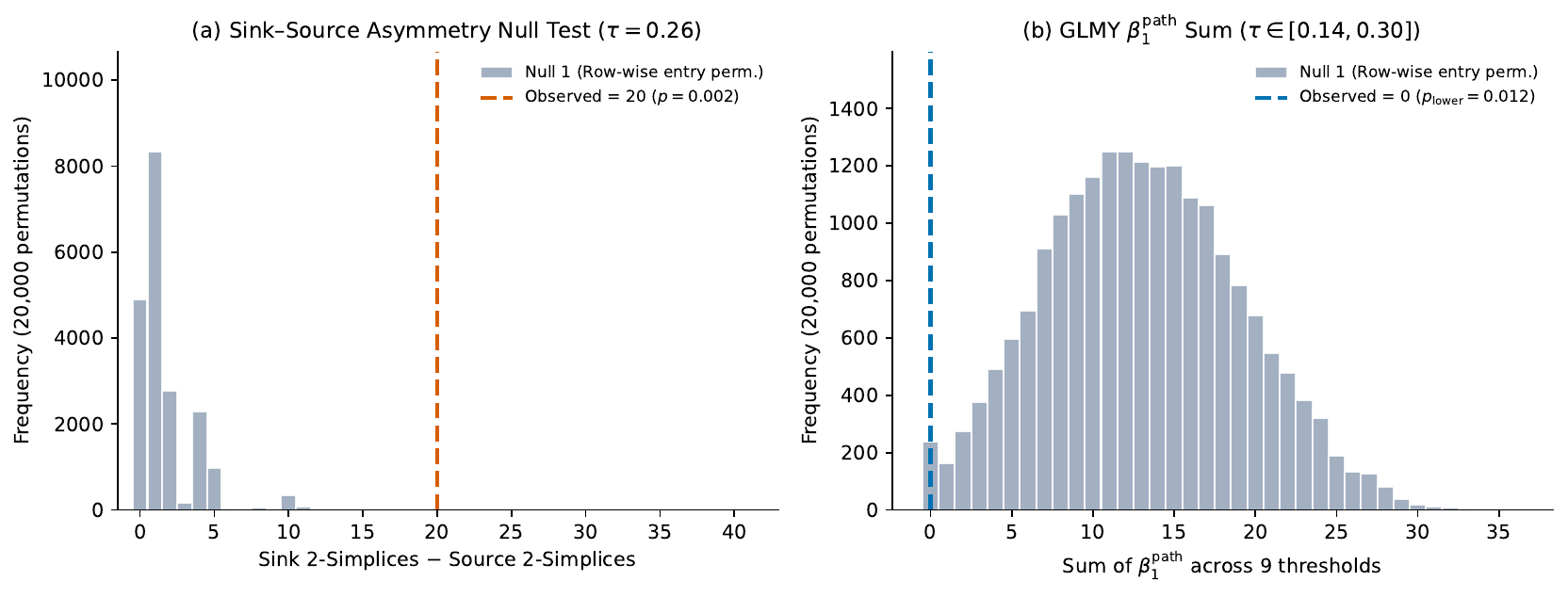}
  \caption{Row-wise entry permutation null distributions ($B = 20{,}000$). (a) Sink--source asymmetry at $\tau = 0.26$ ($p = 0.002$). (b) Regular GLMY $\beta_1^{\mathrm{path}}$ sum across $\tau \in [0.14, 0.30]$ (observed $= 0$, lower-tail $p = 0.012$).}
  \label{fig:null}
\end{figure}

\mbox{Table~\ref{tab:null}} and \mbox{Fig.~\ref{fig:null}} summarize the null model tests. Under row-wise entry permutation (which fixes out-weights), sink--source asymmetry is highly significant ($p = 0.003$ at $\tau = 0.20$; $p = 0.002$ at $\tau = 0.26$), confirming significant sink concentration. Relative to this row-permutation null, the GLMY $\beta_1^{\mathrm{path}}$ sum ($0$ vs.\ $13.070$, lower-tail $p = 0.012$) is significantly lower and the $\beta_1^{dFl}$ sum ($15$ vs.\ $29.441$, lower-tail $p = 0.040$, unadjusted) is lower than random expectations over the nine thresholds $[0.14, 0.30]$. This lower cycle sum primarily reflects the strong in-degree concentration onto \texttt{consensus}: once the in- and out-degrees at $\tau = 0.26$ are held fixed, $\beta_1^{\mathrm{path}} = 0$ occurs in $52.4\%$ ($229/437$) of all digraphs with that degree sequence ($p = 0.523$ in rewiring), and the symmetrization blind spot occurs in $32.4\%$ of rewiring replicates (exact $142/437 = 0.325$). These findings clarify that the empirical value of $dFl$ and Dowker complexes lies in representing the specific \texttt{consensus} $\leftrightarrow$ \texttt{plan} loop, the directed 3-cycle \texttt{\{consensus, coregulate, emotion\}}, and \texttt{consensus} sink convergence as topological summaries over the directed graph.

\FloatBarrier
\section{Conclusion}
Integrating Dowker sink/source complexes, directed flag complexes, and regular GLMY path homology with direct cycle enumeration preserves edge orientation and the orientation order of directed triangles in TNA. On \texttt{group\_regulation}, this approach captures structural facets that \texttt{tna::cliques(size = 2)} and \texttt{InStrength} surface only as separate summaries and that max-symmetrization obscures: (1) the persistent reciprocal \texttt{consensus} $\leftrightarrow$ \texttt{plan} loop across $\tau \in [0.20, 0.28]$ ($\beta_1^{dFl} - \beta_1^{\mathrm{sym}} = 1$, with $\beta_1^{\mathrm{sym}} = 0$ for $\tau \in (0.239087, 0.290401]$) and the directed 3-cycle \texttt{\{consensus, coregulate, emotion\}} over $(0.146825, 0.172081]$ (present at $\tau = 0.15$ and $\tau = 0.16$), and (2) the statistically significant convergence of six regulatory behaviors onto \texttt{consensus} ($20$ sink 2-simplices vs.\ $0$ source 2-simplices, $p = 0.002$), accompanied by lower GLMY $\beta_1^{\mathrm{path}}$ sums than row-permutation baselines ($p = 0.012$ over $[0.14, 0.30]$, driven by in-degree concentration).

For learning analytics researchers interpreting collaborative regulation, preserving these directed structures recovers substantive distinctions that undirected symmetrization collapses. Specifically, the reciprocal \texttt{consensus} $\leftrightarrow$ \texttt{plan} loop separates iterative planning--agreement negotiation from one-way progression, the unfilled directed 3-cycle \texttt{\{consensus, coregulate, emotion\}} distinguishes cyclic socio-emotional and co-regulatory feedback from feed-forward transitive sequences, and the 20-versus-0 sink--source asymmetry identifies \texttt{consensus} as a shared convergent hub rather than a divergent initiator. Future work will extend this pipeline as a complementary directed layer alongside existing \texttt{tna} and \texttt{Nestimate} workflows to multi-group comparisons and larger event-log benchmarks.

All R and Python scripts required to reproduce the filtration analyses, null-model tests, and sequence-level bootstrap evaluations are publicly available at \url{https://github.com/igu-lab/directed-transition-topology} and archived on Zenodo (\url{https://zenodo.org/records/24301782}).

\begin{credits}
\subsubsection{\ackname}
This work was supported by JSPS KAKENHI Grant Numbers JP25K00845, JP25K21951, JP26K00528, and JP26K22187.

\subsubsection{\discintname}
The authors have no competing interests to declare that are relevant to the content of this article.
\end{credits}

\bibliographystyle{splncs04}
\bibliography{references}

\end{document}